# Disseminando a Aprendizagem Colaborativa através do Ambiente Canvas

José Solenir L. Figuerêdo, Renato S. Mascarenhas, Roberto A. Bittencourt

UEFS – Universidade Estadual de Feira de Santana
Av. Transnordestina, s/n, Novo Horizonte
Feira de Santana – BA – Brasil – 44036-900

{solenir.figueredo,natofsa93}@gmail.com, roberto@uefs.br

***Abstract.*** *Virtual learning environments are actual solutions that facilitate collaborative learning, both in classroom and distance education. However, such environments are not yet fully disseminated in Brazilian universities. This work reports a learning experience designed as a workshop that aims to popularize the use of the Canvas virtual learning environment in an university. The workshop emphasizes the support to collaborative learning through resources of the environment. Learned lessons include: the value of adequate planning of workshop activities; the potential of the environment regarding various uses by instructors; and the relevance of collaborative learning for instructors.*

***Resumo.*** *Ambientes virtuais de aprendizagem são soluções atuais que permitem facilitar a aprendizagem colaborativa, tanto presencial quando a distância. Entretanto, tais ambientes ainda não são plenamente difundidos nas universidades brasileiras. Este trabalho apresenta uma experiência de ensino-aprendizagem na forma de oficina com o objetivo de difundir a utilização do ambiente de aprendizagem Canvas em uma universidade, enfatizando o suporte à aprendizagem colaborativa que a plataforma oferece. As principais lições aprendidas foram: o valor do planejamento adequado das atividades da oficina, o potencial do ambiente em relação a formas variadas de utilização pelos professores e, finalmente, a relevância da aprendizagem colaborativa para os professores.*

## 1. Introdução

Aprendizagem colaborativa é um termo genérico para designar uma variedade de abordagens educacionais que envolvem esforço intelectual conjunto de estudantes e professores [Smith and MacGregor 1992]. Para ocorrer uma contribuição positiva ao ambiente educacional, espera-se que formas particulares de interação ocorram entre os envolvidos no processo [Dillenbourg 1999].

Aprendizagem colaborativa requer um ambiente diferente do tradicional [Fuks and Pimentel 2011], onde aluno e professor já não desempenham papeis pretéritos, mas sim novas conotações. Os alunos são ativos e responsáveis por sua própria aprendizagem, sendo o professor apenas um intermediador de práticas pedagógicas que apoiem a formação dos estudantes. Atrelado ao conceito de aprendizagem colaborativa, surge na década de 1990 o conceito de aprendizagem colaborativa apoiada por computador (do inglês, CSCL), um ramo emergente das ciências de aprendizagem que estuda como as pessoas podem aprender juntos com a ajuda de computadores [Stahl et al. 2006].

Existem atualmente diversas ferramentas e plataformas computacionais que dão suporte à aprendizagem colaborativa. Dentre estas, é necessário destacar os ambientes virtuais de aprendizagem, também chamados de sistemas de gerência de aprendizagem. Estes ambientes integram diversos recursos de apoio a aprendizagem, com ênfase na colaboração. Há inúmeros ambientes virtuais de aprendizagem, alguns deles bastante populares, tanto os de código aberto, como o Moodle[1] e Sakai[2], como os proprietários, como Blackboard[3] e Canvas[4].

Entretanto, apesar de existirem inúmeros ambientes de aprendizagem, seu uso não é tão difundido no Brasil, muitas vezes pelo fato de os professores não conhecerem a sua existência ou por não saberem como utilizá-los. Ademais, mesmo havendo algum uso, muitas vezes, as diversas possibilidades que as plataformas oferecem não são aproveitadas em sua plenitude.

O presente trabalho parte deste contexto e visa difundir o ambiente de aprendizagem Canvas entre professores e estudantes da Universidade Estadual de Feira de Santana. Para tanto, desenvolvemos e realizamos uma oficina sobre aprendizagem colaborativa com o ambiente Canvas, oferecida por estudantes do curso de Engenharia de Computação para um público de professores e estudantes da universidade. Esta oficina teve duração de oito horas, e permitiu capacitar, em nível introdutório, um grupo piloto de professores e estudantes, além de oferecer feedback para nosso grupo de pesquisa e extensão melhorar ofertas futuras da oficina.

A oficina realizada levou a algumas lições importantes: os tutores, todos estudantes de graduação, perceberam o valor do planejamento prévio das atividades para garantir o êxito da oficina; os participantes da oficina perceberam e realçaram a variedade de aplicações que podem dar ao ambiente Canvas; perceberam também o potencial da aprendizagem colaborativa na educação, especialmente quando facilitada pelo computador.

O artigo está organizando em seis seções, onde são descritos todos os passos compreendidos nesta pesquisa. A Seção 2 trata da fundamentação teórica, onde são descritos os conceitos que nortearam este trabalho. A Seção 3 explicita a metodologia utilizada e, em seguida, a Seção 4 descreve os resultados obtidos. Na Seção 5, apresentamos em detalhes as lições aprendidas e encerramos com as conclusões e perspectivas de trabalhos futuros na Seção 6.

## 2. Fundamentação Teórica

Nesta seção, descrevemos alguns conceitos importantes relacionados ao desenvolvimento deste trabalho. Abordamos ainda alguns trabalhos relacionados ao presente artigo.

### 2.1. Sistemas Colaborativos

Sistemas colaborativos, termo utilizado largamente no ambiente acadêmico brasileiro, está relacionado a outros dois termos em inglês: groupware e trabalho colaborativo apoiado por computador, cuja sigla em inglês é CSCW [Pimentel and Fuks 2011]. Groupware é um sistema computacional que dá suporte a grupos de pessoas envolvidas em uma tarefa ou objetivo em comum, e que fornece uma interface para um ambiente compartilhado [Ellis et al. 1991]. CSCW, por sua vez, é uma área de investigação que, além de estudar sistemas computacionais que dão suporte a atividades em grupo, também se preocupa com os efeitos psicológicos e sociais que envolvem a colaboração [Pimentel and Fuks 2011]. Estes dois termos surgiram na década de 1980, a partir do interesse comum entre desenvolvedores de produtos e pesquisadores de diversos campos [Grudin 1994].

---

[1] http://www.moodle.org/
[2] https://sakaiproject.org/
[3] http://www.blackboard.com/
[4] http://www.canvaslms.com/

Dentro da categoria de sistemas colaborativos, existem sistemas utilizados no apoio à aprendizagem colaborativa. Na aprendizagem colaborativa, estudantes trabalham ativamente em grupo, construindo o conhecimento a partir da interatividade e da experiência do contexto [Smith and MacGregor 1992]. Um estudante não é apenas responsável pelo seu próprio aprendizado, mas também pelo aprendizado do outro. A aprendizagem colaborativa exige níveis de cognição diferentes da aprendizagem individual, sendo a aprendizagem um processo social no qual os indivíduos constroem seus conhecimentos através da interação com o meio e com as demais pessoas envolvidas no processo [Vygotski 2007]. Neste contexto, a interatividade é central no processo da aprendizagem, e pode ser potencializada pelo apoio de recursos computacionais específicos para aprendizagem colaborativa.

## 2.2. Trabalhos relacionados

Há diversos relatos de experiência do uso de ambientes virtuais de aprendizagem para a aprendizagem colaborativa. Aqui apresentamos alguns trabalhos que mais se aproximam de nossa experiência, como o uso do ambiente Canvas na escola básica [Barros 2013], de colaboração através de blogs [Brito et al. 2009], e de solução de limitações tecnológicas institucionais através de computação em nuvem [Kusen and Hoic-Bozic 2014].

A Escola Básica Integrada da Quinta do Conde, em Portugal, utilizou a plataforma Canvas durante o ano letivo de 2012/2013 para analisar o seu potencial pedagógico em um curso de Educação e Formação de Adultos de Nível Secundário (EFA-NS) [Barros 2013]. Os resultados apontaram para a adequação da plataforma à formação dos estudantes. Já o nosso trabalho utiliza a plataforma Canvas para formar professores potenciais usuários deste ambiente, para utilizá-los em cursos universitários.

O uso de ferramentas de criação de blogs para discutir os princípios da ética em informática foi investigado em outro trabalho, e objetivou investigar a aceitação por parte dos estudantes de ferramentas de blog no apoio à interação em aulas presenciais [Brito et al. 2009]. Verificou-se que, com a utilização dos blogs, as discussões em aula foram enriquecidas, favorecendo a aprendizagem dos estudantes. Blogs são uma ferramenta para aprendizagem colaborativa, enquanto que o nosso trabalho utiliza uma plataforma com várias ferramentas, mas também procura investigar a aceitação da plataforma por seus potenciais usuários.

Outro trabalho relata a migração para um sistema de aprendizagem baseado em computação em nuvem na universidade de Rijeci, na Croácia, como forma de solucionar os problemas de baixo orçamento que o ensino superior enfrenta [Kusen and Hoic-Bozic 2014]. Os resultados sugerem que este tipo de tecnologia contribui de forma satisfatória para o ensino superior, ao atacar restrições financeiras e oferecer uma alternativa para soluções que demandam hardware e infraestrutura mais caros e até mesmo software proprietário. Em nosso caso, a instituição não precisa manter conexões dedicadas nem hardware especializado ao utilizar a plataforma Canvas, embora há a desvantagem de manter os dados de atividades da universidade em uma empresa.

## 3. Metodologia

Nesta seção, descrevemos o objetivo deste trabalho, o ambiente virtual de aprendizagem utilizado, o planejamento da oficina e os procedimentos de coleta e análise de dados.

### 3.1. Objetivos

O presente estudo tem como objetivo geral investigar como ambientes de aprendizagem colaborativa podem contribuir no processo de ensino-aprendizagem em instituições de ensino superior. Um objetivo secundário consiste em disseminar ambientes de aprendizagem colaborativa entre o corpo docente de instituições de ensino superior, popularizando a utilização destes ambientes no espaço acadêmico.

### 3.2. Canvas

Canvas é um ambiente virtual de aprendizagem em nuvem [Canvas 2015a]. Uma outra denominação, dada pelos proprietários da plataforma Canvas, é o de sistema de gerência de aprendizagem (LMS, da sigla em inglês). Um sistema de gerência de aprendizagem é um ambiente virtual de aprendizagem que disponibiliza um conjunto de recursos síncronos e assíncronos que dão suporte ao processo de ensino-aprendizagem [Barros 2013]. Atualmente, existem no mercado diversos sistemas de gerência de aprendizagem, alguns proprietários e outros open source. Apesar de ser um software proprietário, o ambiente Canvas permite o uso gratuito por instituições educacionais. Além da diversidade de recursos e de uma interface amigável no estilo Web 2.0, escolhemos a plataforma Canvas para este trabalho por ser uma plataforma em nuvem, não sendo necessário instalá-lo em um servidor local, o que facilitará a disseminação futura da abordagem proposta em várias instituições de ensino, inclusive na educação básica [Canvas 2015b]. O ambiente Canvas foi lançado pela empresa Instructure no ano de 2011 e continua em permanente evolução [Canvas 2015a]. Dentre os recursos disponíveis, é possível criar portfólios, acessar o ambiente através de dispositivos móveis, utilizar um módulo para análise de aprendizagem, criar avaliações e classificações, além de utilizar diversas ferramentas multimídia, de gestão de resultados da aprendizagem e de apoio à aprendizagem colaborativa, tanto na própria plataforma como pela interoperabilidade com aplicações de terceiros.

### 3.3. Planejamento da Oficina

A oficina de aprendizagem colaborativa com Canvas objetivou introduzir o ambiente Canvas para professores da Universidade Estadual de Feira de Santana (UEFS). Os principais recursos do ambiente foram apresentados na forma de um tutorial, onde os participantes praticavam a criação de cursos no ambiente.

O público convidado foi de professores da instituição, mas acabou contando também com estudantes de licenciatura, interessados na plataforma. Os tutores da oficina intercalaram conceitos teóricos e práticos, visando elucidar como um dado recurso poderia ser aplicado em sala de aula, e como poderia contribuir na colaboração entre os estudantes.

Na Tabela 1, pode-se observar a descrição geral da oficina, com objetivos, metodologia empregada, conteúdo abordado, além de outras informações.

**Tabela 1. Descrição Geral da Oficina.**

| Oficina de Aprendizagem Colaborativa com Canvas ||
|---|---|
| **Objetivos** | Ser capaz de utilizar os principais recursos do ambiente Canvas para a criação de cursos de apoio a disciplinas presenciais ou a distância de modo a melhorar a interatividade entre professores e estudantes. |
| **Metodologia** | Tutorial apoiado por projetor multimídia e laboratório de informática, combinando exposições com atividades práticas. |
| **Conteúdo** | Criação de contas e cursos. Aspectos gerais e configuração de cursos. Cadastro e manutenção de pessoas, grupos e discussões. Eventos e Anúncios. Tarefas, testes e notas. Cronograma, páginas e wikis. Módulos. Conferência e aplicativos. |
| **Tutores** | Dois estudantes do curso de Engenharia de Computação. |
| **Local** | Laboratório de Informática do PPGSC-UEFS. |
| **Participantes** | 6 professores e estudantes de graduação da UEFS. |
| **Período** | 1 semana. |
| **Carga Horária** | 8 horas, divididas em 2 sessões de 4 horas. |

Vale asseverar que se procurou trabalhar os conteúdos de maneira gradual, realçando os elementos de apoio à aprendizagem colaborativa, e sua aplicabilidade na educação. Como pode ser observado na Tabela 2, o planejamento seguiu uma ordem de complexidade crescente, partindo dos recursos mais fáceis que a plataforma oferece, como a criação de cursos, páginas, grupos, tarefas, anúncios, eventos, e findando com tópicos com um maior grau de dificuldade, como a criação de testes, atribuição de notas, criação de wikis, programa, módulos, conferência e colaboração, dentre outros.

**Tabela 2. Planejamento resumido da oficina.**

| Duração | Objetivos | Atividade | Conteúdo |
|---|---|---|---|
| 2 horas | Ter o primeiro contato com o ambiente, acessando a ferramenta para criar cursos e adicionar pessoas. | Tutorial combinando exposição e práticas. | Introdução. Criação de conta e curso. Página inicial de um curso. Adição de pessoas. |
| 2 horas | Ser capaz de criar grupos dentro do curso, criar tarefas e eventos e fazer anúncios dentro da ferramenta. | Tutorial combinando exposição e práticas. Grupos são formados para simular classes de um curso. | Eventos e Anúncios. Grupos. Tarefas. |
| 2 horas | Ser capaz de avaliar os alunos dos cursos com a criação de testes e a atribuição de notas, além de criar discussões, páginas e inserir arquivos num curso. | Tutorial combinando exposição e práticas. | Testes e Notas. Discussão. Páginas. Arquivos. |
| 2 horas | Ser capaz de dividir o curso em módulos e criar um programa de curso, além de fazer conferências e edição de arquivos de forma colaborativa, utilizar aplicativos externos e fazer configurações mais avançadas. | Tutorial combinando exposição e práticas. Conferência com participantes. Criação colaborativa de arquivos. | Programa. Módulos. Conferência. Colaboração. Configurações. Aplicativos. |

### 3.4. Coleta e análise de dados

Ao término de cada sessão da oficina, os dois tutores escreveram diários de bordo, registrando suas impressões sobre a sessão e descrições do andamento, objetivando auxiliar na análise. Ao final da oficina, os participantes responderam a um questionário subjetivo, utilizando o próprio ambiente de aprendizagem. O questionário tratava de questões referentes à experiência da oficina, o ambiente Canvas e a metodologia empregada. O anonimato dos participantes foi preservado nos questionários.

Os dados coletados foram analisados utilizando através de procedimentos qualitativos preliminares. Como se tratou da primeira oferta desta oficina, optamos por uma análise qualitativa mais simplificada. Ofertas futuras utilizarão entrevistas mais detalhadas e observação participante na coleta de dados, e codificação e análise de conteúdo para a análise dos resultados.

## 4. Resultados

Como já fora explanado, a oficina utilizando a plataforma Canvas atendeu a seis professores universitários e dois estudantes de licenciatura. As características deste público facilitaram o andamento da oficina, tornando-a bastante produtiva durante quase todo o tempo.

A oficina foi iniciada com uma introdução ao ambiente Canvas, descrevendo seu histórico, suas vantagens, seu potencial na educação, e seu uso atual pelas universidades. De maneira superficial, introduziu-se o tema de sistemas colaborativos, para que os participantes se situassem de maneira mais ampla sobre os objetivos da oficina. Os participantes que estavam presentes demonstravam atenção e interesse pela apresentação. Terminada esta introdução, foi iniciada a parte prática da oficina.

O ambiente Canvas possui diversos recursos que contribuem para a aprendizagem. Cada recurso do Canvas foi demonstrado, sendo explicado como poderia ser utilizado em sala de aula e quais as suas contribuições ao processo de aprendizagem. Em seguida, os participantes eram convidados a utilizarem estes recursos, após os tutores explicarem passo a passo como fazê-lo. A cada recurso utilizado, era notável o entusiasmo e a satisfação dos participantes com o ambiente. Eles falavam entre si frases do tipo: *"Nossa! A ferramenta é ótima."*, *"Pode-se aplicar em diversos contextos"* ou *"Eles pensaram em tudo"*.

Para permitir que os participantes trabalhassem tanto na visão de professor como na visão de estudante, a turma foi dividida em trios, e foi solicitado que cada participante cadastrasse os outros dois em um curso criado previamente criado por eles no ambiente. Desta forma, quando eles criavam algum recurso disponibilizado pela plataforma, eles podiam observar a visão do aluno e a do professor. Com isto, pôde-se observar a aplicabilidade da ferramenta com boa aproximação com a realidade.

O ambiente dispõe de diversos recursos que viabilizam o trabalho em grupo, tais como discussões, tarefas, testes, conferência e colaboração. Estes recursos que envolviam grupos foram os que os participantes mais apreciaram, pois eles viam nestes recursos uma oportunidade de viabilizar a aprendizagem colaborativa com os alunos deles.

Em ferramentas de aprendizagem online, a avaliação é um dos itens mais delicados. Muitas vezes, o ambiente não fornece um suporte tão eficiente ou variado. Para os participantes da oficina, entretanto, esta percepção foi diferente com o Canvas. O ambiente contém recursos que dão suporte tanto à avaliação somativa quanto à avaliação formativa. Os participantes da oficina puderam verificar estes dois tipos de avaliação. Para isso, fez-se uso dos recursos discussão, tarefa e testes, mostrando as duas vertentes de avaliação. Os participantes puderam analisar por meio de gráficos o grau de interação dos estudantes em um curso, além de visualizar a evolução dos alunos nas atividades realizadas no ambiente.

Os participantes se mostraram sempre interessados no ambiente, e viram nele um grande potencial. Também tiveram muita facilidade em utilizar os recursos do ambiente. Eles mesmos relataram que o ambiente é muito fácil de ser utilizado, e que pode contribuir muito para o trabalho desenvolvido por eles.

### 4.1. Questionário aplicado

A aplicação de um questionário subjetivo foi relevante pois pôde fornecer informações relativas à percepção dos participantes sobre a utilidade do ambiente e sobre a metodologia empregada pelos tutores na oficina. Ademais, o questionário foi respondido utilizando o próprio ambiente, o que foi interessante, pois os participantes tinham acabado de passar pelo aprendizado da ferramenta, e já estavam utilizando-a para uma atividade real e útil.

Numa das respostas, uma participante salientou que *"o assunto abordado foi extremamente interessante e a possibilidade da sua aplicação no processo ensino-aprendizagem é real. A ferramenta Canvas atende a todas as necessidades de um professor. Agregando não somente o processo de ensino, mas também a interlocução professor-aluno e as discussões necessárias. Fiquei muito satisfeita com a oportunidade e o convite para assistir a este curso"*. Percebe-se, da resposta desta participante, o anseio por um ambiente de aprendizagem inovador e eficaz.

Outro participante relatou: *"Achei tudo muito interessante, em especial o uso de aplicativos para a colaboração e o item conferências"*. Nota-se, nesta fala, o quanto é necessário haver recursos que proporcionem a colaboração entre os alunos. Em geral, acerca da metodologia empregada, os participantes informaram que foi bem satisfatória, pela abordagem prática adotada de realização de atividades relacionadas à ferramenta.

Uma das participantes sugeriu ainda: *"Fazer um curso de Canvas usando a própria ferramenta e deixar disponível para os alunos que cursaram este curso, de modo que eles possam disseminar mais rápido junto aos seus alunos de mestrado e iniciação científica"*.

### 5. Lições aprendidas

Como este trabalho é essencialmente um relato de experiência na forma de um estudo de caso descritivo, vale a pena registrar algumas lições importantes aprendidas com a realização da oficina de aprendizagem colaborativa.

*Planejamento.* Uma das primeiras lições aprendidas é sobre o valor de um planejamento adequado. Durante a oficina, pôde-se observar que tudo ocorreu conforme o planejado. A ordem temporal dos temas também proporcionou segurança aos participantes, relatada por eles no questionário aplicado ao final da oficina.

*Aplicações da ferramenta.* Durante a oficina, foi possível observar as várias aplicações do ambiente sugeridas pelos participantes. Comentários como *"Nossa! Vou utilizar com os meus orientandos para criar a turma do estágio no Canvas"* ou *"Posso criar trabalhos lúdicos com professores e alunos de sétima série"* eram frequentes. Isto nos revelou mais do que havíamos sugerido, fazendo-nos perceber várias novas possibilidades de aplicação e novas ideias para as próximas oficinas que pretenemos aplicar no decorrer do ano presente.

*Aprendizagem Colaborativa.* A partir desta experiência, notou-se, que a aprendizagem colaborativa é de fato um tema de grande importância para os professores e estudantes, podendo ser aplicada tanto em ambientes presenciais como apoiados por ambientes virtuais. Percebe-se a necessidade de aplicar este conceito mais frequentemente: os professores da oficina relataram que precisavam de ferramentas deste tipo para tornar suas metodologias de ensino mais dinâmicas e proporcionar a aprendizagem em grupo de forma mais interativa.

*Tempo.* O tempo foi um dos poucos pontos negativos, pois, apesar do planejamento detalhado, sentiu-se a necessidade de mais tempo para explicação de alguns tópicos. Outros tópicos acabaram levando menos tempo por serem mais simples, ou pela desenvoltura dos participantes. Com isso, percebemos que é necessário reaplicar a oficina com mais eficácia. De todo modo, entendemos que o tempo curto de oito horas é o adequado para ter alguma profundidade e, ao mesmo tempo, permitir que professores bastante ocupados possam ter acesso a este aprendizado.

## 6. Conclusões

Este artigo relatou a experiência de uma oficina que visou promover a aprendizagem colaborativa através do ambiente virtual de aprendizagem Canvas. Este ambiente, como outros ambientes virtuais, proporciona a aprendizagem colaborativa de forma online, permitindo maior interação entre os estudantes e entre estudantes e o professor, encurtando as distâncias entre eles. Com isso, o ambiente de aprendizagem não fica restrito apenas ao ambiente escolar ou acadêmico. O estudante agora faz uso de outros espaços, tendo acesso aos recursos de maneira online, onde quer que esteja.

A realização desta oficina foi um primeiro passo necessário para popularizar os ambientes virtuais de aprendizagem na Universidade Estadual de Feira de Santana. Acreditamos que, de maneira indireta, ocorrerá a disseminação destes ambientes de aprendizagem nas escolas, pois esta universidade possui diversos cursos de licenciatura, e, ao aprender através de um ambiente, é possível que os egressos dos cursos de licenciatura levem estas tecnologias ao ambiente escolar, e fortalecendo a utilização da aprendizagem colaborativa neste ambiente.

Em trabalhos futuros, pretendemos realizar outras oficinas com objetivos similares, mas fazendo uso de mecanismos de pesquisa mais elaborados, como questionários quantitativos, entrevistas, observação participante, usando uma abordagem de pesquisa de estudo de caso quali-quantitativo. Pretendemos ainda disponibilizar um tutorial passo-a-passo sobre os recursos do ambiente Canvas, já que a disponibilidade de material em língua portuguesa sobre este ambiente é escasso.

## Referências

Barros, C. A. da C. (2013). Proposta de curso no âmbito dos Cursos de Educação e Formação de Adultos de Nível Secundário. Utilização da plataforma Canvas na formação escolar e profissional. Universidade de Lisboa.

Brito, J. A., Souza, F. V, Silva, J. A. and Gomes, A. S. (2009). O Blog como Ferramenta de Aprendizagem Colaborativa: uma experiência em um curso de formação técnica. In XX Simpósio Brasileiro de Informática na Educação (SBIE), Florianópolis, SC.

Canvas (2015a). Guide Canvas. https://community.canvaslms.com/community/answers/guides.

Canvas (2015b). Compare Canvas. http://www.canvaslms.com/higher-education/features.

Dillenbourg, P. (1999). What do you mean by ' collaborative learning  "? Collaborative learning Cognitive and computational approaches, v. 1, n. 6, p. 1–15.

Ellis, C. A., Gibbs, S. J. and Rein, G. (1991). Groupware: some issues and experiences. Communications of the ACM. ACM.

Fuks, H. and Pimentel, M. (2011). Aprendizagem colaborativa com suporte computacional. Sistemas Colaborativos. p. 135–157.

Grudin, J. (1994). Computer-supported cooperative work: History and focus. Computer,

Kusen, E. and Hoic-Bozic, N. (2014). Use of New Technology in Higher Education: A Migration to a Cloud-Based Learning Platform. In The 2nd International Workshop on Learning Technology for Education in Cloud. . Springer.

Pimentel, M. and Fuks, H. (2011). Teorias e modelos de colaboração. Sistemas Colaborativos. p. 16 – 34.

Smith, B. L. and MacGregor, J. T. (1992). What is collaborative learning?

Stahl, G., Koschmann, T. and Suthers, D. (2006). Computer-supported collaborative learning: An historical perspective. Cambridge handbook of the learning sciences, p. 409–426.

Vygotski, L. S. (2007). A formação social da mente. São Paulo: Martins Fontes.